\documentclass[12pt,preprint]{aastex}

\usepackage{epsfig}
\usepackage{amsmath}
\usepackage{natbib}
\usepackage{verbatim}
\usepackage{rotating}
\usepackage{lscape}

\newcommand{\etal}{{\rm et al.}}
\newcommand{\etals}{{\rm et al.} }

\newcommand{\hi}{H \sc{i}\rm{ }}
\newcommand{\hins}{H \sc{i}\rm}

\slugcomment{ApJ Letters, {\em in press}, 10 May 2011}
\shorttitle{A Molecular Arm in the Far Outer Galaxy}
\shortauthors{Dame and Thaddeus}

\begin{document}

\title{A Molecular Spiral Arm in the Far Outer Galaxy}
\author{T. M.\ Dame and P.\ Thaddeus}
\affil{Harvard-Smithsonian Center for Astrophysics, 60 Garden Street, Cambridge MA 02138 
 \\ tdame@cfa.harvard.edu, pthaddeus@cfa.harvard.edu}

\begin{abstract}
We have identified a spiral arm lying beyond the Outer Arm in the first Galactic quadrant $\sim$15 kpc from the Galactic center. After tracing the arm in existing 21 cm surveys, we searched for molecular gas using the CfA 1.2 meter telescope and detected CO at 10 of 220 positions. The detections are distributed along the arm from $l = 13^{\circ}$, $v = -21$ km s$^{-1}$ to $l = 55^{\circ}$, $v = -84$ km s$^{-1}$ and coincide with most of the main \hi concentrations. One of the detections was fully mapped to reveal a large molecular cloud with a radius of 47 pc and a molecular mass of $\sim$50,000 M$_\odot$. At a mean distance of 21 kpc, the molecular gas in this arm is the most distant yet detected in the Milky Way. The new arm appears to be the continuation of the Scutum--Centaurus Arm in the outer Galaxy, as a symmetric counterpart of the nearby Perseus Arm.

\end{abstract}



\section*{}
Although many models of our Galaxy's spiral structure have been proposed during the past 50 years, unambiguous evidence for the type of twofold symmetry expected of a barred spiral galaxy such as ours has remained elusive. Such evidence requires tracing spiral arms on the far side of the Galaxy, where those beyond the solar circle are both deficient in molecular gas and extremely distant, and those within are largely masked by near-side emission at the same velocity. An important exception within the solar circle is the Far 3-kpc Arm \citep{Dame08}; its clear symmetry with respect to its near-side counterpart is apparent only because of the large outward motions of both arms. Here we describe the detection of another far-side molecular arm, one well beyond the solar circle in the first Galactic quadrant. Although the arm is still only partially observed in CO, it is evident in \hi as well, and its existence as a coherent Galactic structure of large size and substantial mass seems beyond doubt.  It is most likely the extension of the Scutum--Centaurus (Sct-Cen) Arm in the outer Galaxy and the symmetric counterpart of the Perseus Arm, which passes within a few kpc of the Sun.
	 
The new arm was found as a result of attempts to follow the Sct--Cen Arm past its tangent near $l = 309^{\circ}$ and into the fourth quadrant.  For reasons discussed below this attempt failed, but even farther from its tangent in the distant first quadrant the arm, to our surprise, emerged clearly, first in the LAB 21 cm survey \citep{Kalberla05} and then, guided by poorly resolved clumps in the \hins, in new CO observations with the CfA 1.2 m telescope. The present Letter is largely devoted to the molecular arm, which, because of the more than fourfold higher angular resolution in the CO data, should define the arm more sharply than the \hins. 
	 	
The new arm was largely overlooked in existing 21 cm surveys probably because it lies mainly out of the Galactic plane, its Galactic latitude steadily increasing with longitude as it follows the warp in the distant outer Galaxy.  As Figure 1a shows, in the Galactic plane in the first quadrant the only prominent \hi spiral feature in the outer Galaxy (i.e., at negative velocities) is the well-known Outer Arm, a feature also well traced by CO (see, e.g., Fig. 4 below). However, at $3^{\circ}$ above the plane (Fig. 1b) one sees instead the new arm as a prominent linear feature running roughly parallel to the locus of the Outer Arm but shifted by $\sim$30 km s$^{-1}$ to more negative velocities. Even with the velocity shift, in longitude-velocity maps at positive latitudes such as that in Figure 1b, the arm can easily be mistaken for the Outer Arm.  To our knowledge, the only study that traced both the Outer Arm and the new arm as separate features in longitude and velocity is that of Weaver (1974; Fig. 7), although neither feature was explicitly discussed.  More recently, Binney \& Merrifield (1998; Fig. 9.21) noticed the short segment of this arm that is visible at $b = 0^{\circ}$, but, again, made no comment.
	 
One might wonder whether the feature at $b = 3^{\circ}$ in Figure 1b is in fact the Outer Arm which is subject to a large "rolling motion" of the sort discussed by \citet{Feitzinger86} and others, its velocity shifting more negative with latitude. However, this possibility is readily excluded. Both arms are seen distinctly at many latitudes intermediate between those shown in Figure 1, and latitude-velocity maps at many longitudes (e.g., Fig. 2) likewise show two distinct emission features, not the single inclined feature expected of rolling motions.  Furthermore, the roughly linear locus of the new arm in longitude and velocity extrapolates very close to zero velocity at zero longitude (Fig. 1b, as well as Figs. 3b and 4 below)---the clear signature of an arm lying well beyond the solar circle at zero longitude. This is in marked contrast to the Outer Arm, which passes inside the solar circle (to positive velocities) at $l < 20^{\circ}$ (Fig. 1a).

\begin{figure}
\centering
\epsscale{1.}
\plotone{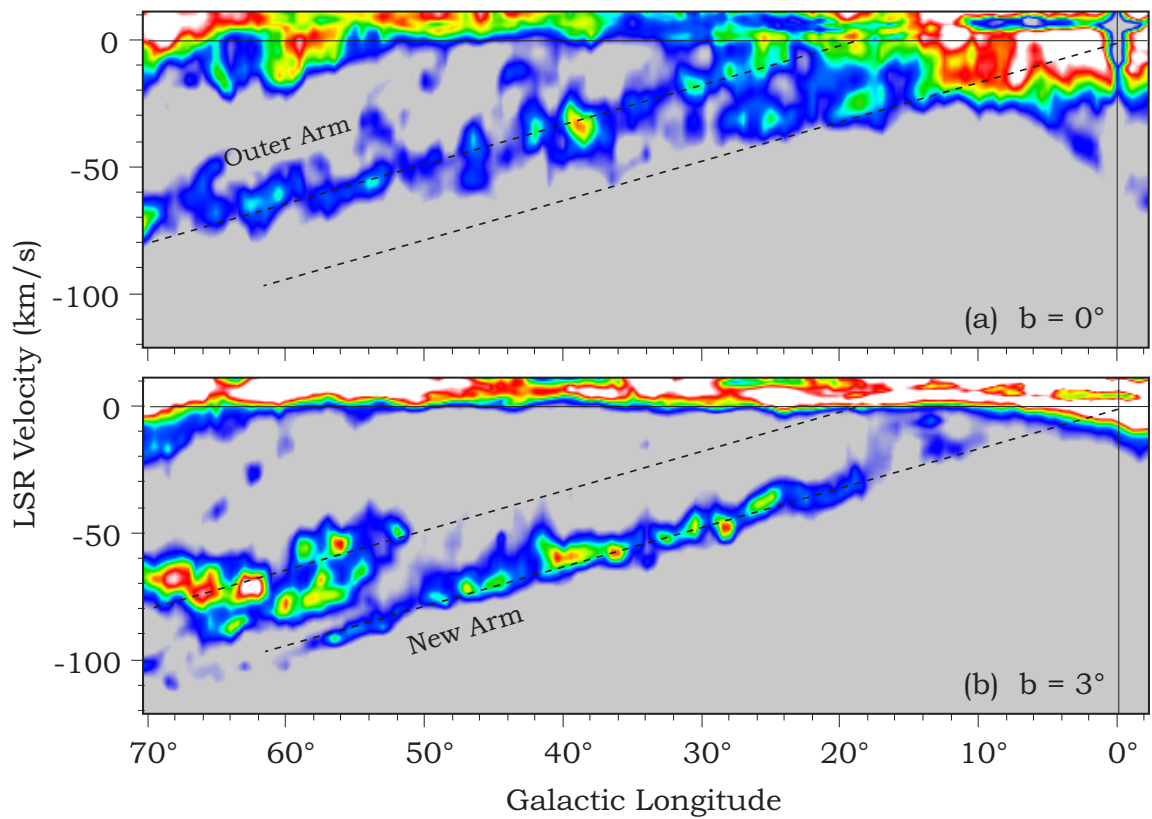}
\caption{Longitude-velocity diagrams of 21 cm emission at (a) $b = 0^{\circ}$ and (b) $b = +3^{\circ}$, from the LAB \hi survey \citep{Kalberla05}.  The color palette from light blue to white corresponds to the 
intensity range 30--110 K in (a) and from 15--55 K in (b). The approximate locus of the Outer Arm is marked by the upper dashed line in both panels and the new arm by the lower dashed line in both. 
 }
\end{figure}

Figure 3a is a spatial map of the new arm in \hins, obtained by integrating the LAB survey over a window that follows the arm in velocity, and, similarly, Figure 3b is a longitude-velocity map that follows the arm in latitude. The arm is well defined over $\sim$$60^{\circ}$ of Galactic longitude with a thickness that varies slightly, from 1.1 kpc (FWHM) at low longitudes to 1.6 kpc at high, assuming a mean distance of 21 kpc. The arm's vertical offset varies much more, from lying in the plane at low longitudes to lying $\sim$1.5 kpc above it at high longitudes. These variations are consistent with the general warping and flaring of the outer \hi disk in the first quadrant, as measured by Levine, Blitz, \& Heiles (2006; Figs. 2 \& 4).  Assuming the IAU values for R$_\odot$ and V$_\odot$  and a flat rotation curve beyond the solar circle, the arm's Galactic radius varies from $\sim$14 kpc at $l = 10^{\circ}$ to 17 kpc at $60^{\circ}$, its distance from the Sun decreasing slightly over the same range, from 23 to 20 kpc.
	 
The positions in the new arm at which we have so far detected CO with the CfA 1.2 meter telescope are marked by the yellow dots in both panels of Figure 3, and the data are summarized in Table 1. The telescope is equipped with a sensitive SIS receiver with a single sideband noise temperature of 60 K, yielding total system temperatures that vary with weather and elevation in the range 350-800 K.  Its beamwidth at 115 GHz is 8.4\arcmin, corresponding to 50 pc at 21 kpc.  Observations were frequency switched over 15 MHz at a rate of 1 Hz.  Total integration times per point varied from 20 to 75 minutes, adjusted automatically on the basis of the system temperature to achieve an rms after folding of 0.02 K in each 0.65 km s$^{-1}$ channel. It is worth emphasizing that observations such as these are entirely routine for the telescope, which has successfully carried out more sensitive and more technically difficult surveys during the past two decades (e.g., M31; Dame et al. 1993).

\begin{figure}
\centering
\epsscale{0.7}
\plotone{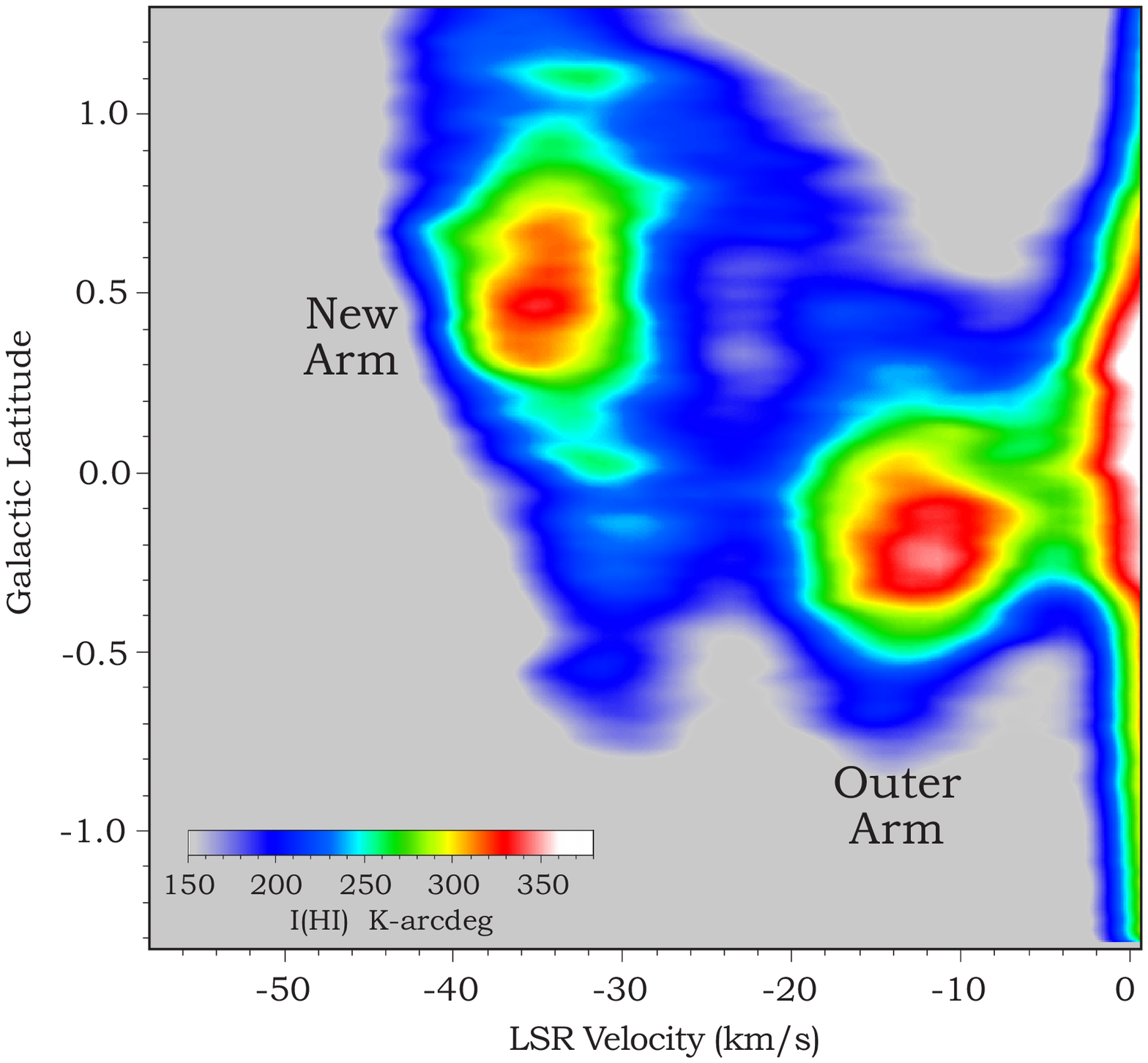}
\caption{
Latitude-velocity diagram of 21 cm emission from the VLA Galactic plane survey (VGPS; Stil et al. 2006), integrated over the longitude range $25^{\circ}-30^{\circ}$.  In this high-resolution (1\arcmin) survey, the Outer Arm and new arm are clearly seen as strong, well-resolved features. Unfortunately, both arms are visible in the VGPS over only a short range of longitude: at $l < 20^{\circ}$ the Outer Arm is confused with local emission at $v > 0$ km s$^{-1}$,  and at $l > 35^{\circ}$ the new arm is mainly above the survey's latitude limit of $1.3^{\circ}$.  
}
\end{figure}

Our search for CO in the arm has so far been guided by the LAB \hi survey (Fig. 3) and our preliminary goal---now almost achieved---was to detect molecular gas in each of the main \hi concentrations along the arm. To date, 220 positions along the arm have been searched, with CO detected in 5\% of them. At a mean distance of $\sim$21 kpc, these are the most distant detections of CO in the Milky Way. No previous CO survey of the region had the necessary combination of sensitivity, latitude coverage, and velocity coverage to detect these clouds. A complete CO survey of the new arm with the CfA telescope is planned, but this will require observation of $\sim$6,400 positions over the course of two observing seasons.

\begin{figure*}
\centering
\epsscale{1.}
\plotone{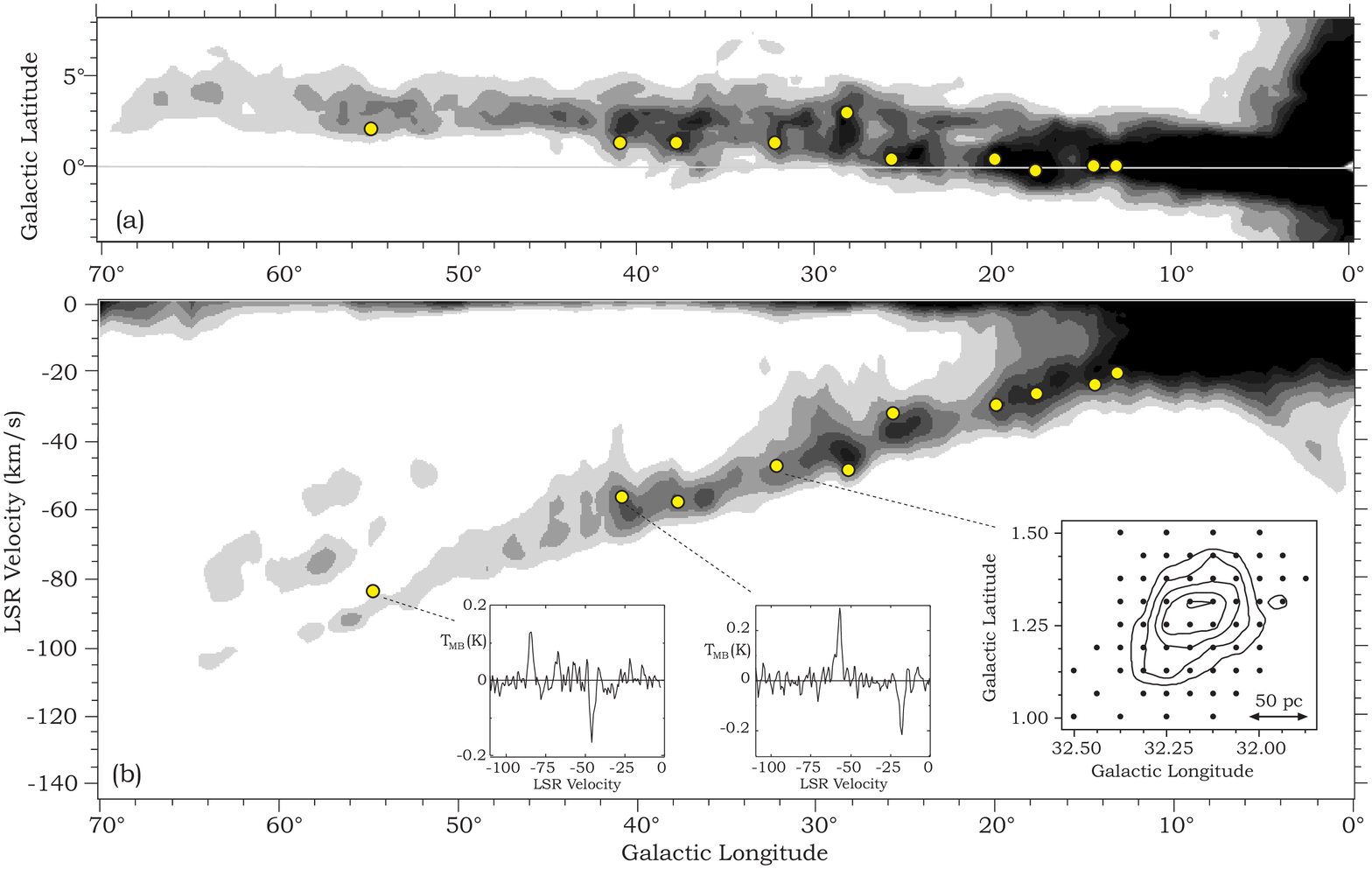}
\caption{ (a) Velocity-integrated 21 cm emission from the new arm, obtained by integrating the LAB survey over a window 14 km s$^{-1}$ wide that follows the arm in velocity. The window is centered on the lower dashed line shown in both panels of Fig. 1, defined by the equation $v = -1.6 * l$. The grayscale levels run from 100 K km s$^{-1}$ ({\em light gray}) to 600 K km s$^{-1}$ ({\em black}).  (b) Longitude-velocity diagram of 21 cm emission from the new arm, obtained by integrating the LAB survey over a window $3.5^{\circ}$ wide that follows the arm in latitude. The window is centered on the line $b = 0.375^{\circ} + 0.075^{\circ} * l$. The grayscale levels run from 40 K-arcdeg ({\em light gray}) to 160 K-arcdeg ({\em black}). The insets show two sample spectra and a velocity-integrated CO map of the one cloud we fully mapped. {\em The spectra are frequency switched and unfolded, so the lines are seen in both phases}, separated by 39 km s$^{-1}$; the LSR velocity scale corresponds to the signal phase (positive lines). In the CO map the contour interval is 0.16 K km s$^{-1}$, starting at 0.16 K km s$^{-1}$, the black dots are the observed positions, spaced every half beamwidth, and the 50 pc arrow assumes a distance of 20.8 kpc (Table 1). In both (a) and (b), yellow dots mark CO detections; the beam of the 1.2 m telescope is five times smaller. 
 }
\end{figure*}

\begin{table}
\caption{CO Detections in the New Arm\label{tbl-1}}
\begin{tabular}{ccccccc}
\tableline\tableline
$l$ & $b$ & T$_{MB}$ & v$_{LSR}$\tablenotemark{a} & $\Delta$v\tablenotemark{b} & W$_{CO}$\tablenotemark{c} & d\tablenotemark{d}  \\


($^{\circ}$) & ($^{\circ}$) & (K) & &  &  & \\

\tableline

13.250  &	0.000 & 0.27 & -20.9 & 3.32 & 0.78 & 22.7\\
14.500 & 0.000	 & 0.32 & -24.4 & 4.72  & 1.31 & 23.3 \\
17.750 & -0.250 & 0.22 & -27.3 & 4.03 & 0.77 & 22.2 \\
20.000 & 0.375 & 0.47 & -30.4 & 3.81 & 1.54 & 21.9  \\
25.750 & 0.375 & 0.11 & -32.2 & 5.03 & 0.48 & 19.9  \\
28.250 & 2.875 & 0.15 & -49.2 & 5.17 & 0.66 & 23.1 \\
32.250 & 1.250 & 0.14 & -47.6 & 6.67 & 0.78 & 20.8 \\
37.750 & 1.250	 & 0.15 & -58.7 & 3.20 & 0.41 & 20.9 \\
40.875 & 1.250	 & 0.23 & -57.0 & 3.24 & 0.66 & 19.4 \\
54.750 & 2.000	 & 0.15 & -84.4 & 2.95 & 0.38 & 19.4 \\
\tableline
\end{tabular}

\tablenotetext{a}{Velocity from Gaussian fit (km s$^{-1}$)}  
\tablenotetext{b}{FWHM linewidth from Gaussian fit (km s$^{-1}$)}  
\tablenotetext{c}{Velocity-integrated line intensity (K km s$^{-1}$)}
\tablenotetext{d}{Kinematic distance (kpc)}
\end{table}


We have fully mapped one of the typical detections near the middle of the arm every half beamwidth and found a large molecular cloud with a radius of 47 pc and molecular mass of at least 2.5 x $10^{4}$ M$_\odot$ (Fig. 3b {\em inset}).  This mass is very likely a lower limit, since it is based on the CO-to-H$_2$ mass conversion factor X measured in the solar neighborhood (1.8 x 10$^{20}$ cm$^{-2}$ K$^{-1}$ km$^{-1}$ s; Dame, Hartmann \& Thaddeus 2001).  A recent study of the X factor in the outer Galaxy based on modeling of the diffuse gamma ray emission observed by the {\em Fermi} satellite \citep{Abdo10} suggests that X rises by a factor of $2-3$ between the solar circle and the 14 kpc radius of this cloud. Such a rise is consistent with the corresponding falloff in metallicity over the same range of radii measured by \citet{Rolleston00}. It would appear then that at least some of the molecular gas in this arm is contained in giant molecular clouds of the sort found elsewhere in the Galaxy.

The new arm is best interpreted as the far northern extension of the Sct--Cen Arm, the main component of the so-called Molecular Ring collection of clouds about half way between the Sun and the Galactic center. As Figure 4 shows, a logarithmic spiral constrained to fit the well-defined tangent directions of the Scutum Arm in the first quadrant and the Centaurus Arm in the fourth extrapolates into the outer first quadrant as a nearly straight line in longitude-velocity space that fits extremely well the new arm in both CO and \hins.  No other known spiral arm can be plausibly extrapolated in this way to account for the new arm.  The Sct--Cen Arm as shown in the inset to Figure 4 matches quite well the schematic sketch of the Galaxy in \citet{Churchwell09}; there the Sct--Cen Arm past its southern tangent was drawn on the basis of nothing more than extrapolation and a desire to show symmetry with the Perseus Arm (Benjamin \& Hurt, priv. comm.).

\begin{figure*}
\centering
\epsscale{1.}
\plotone{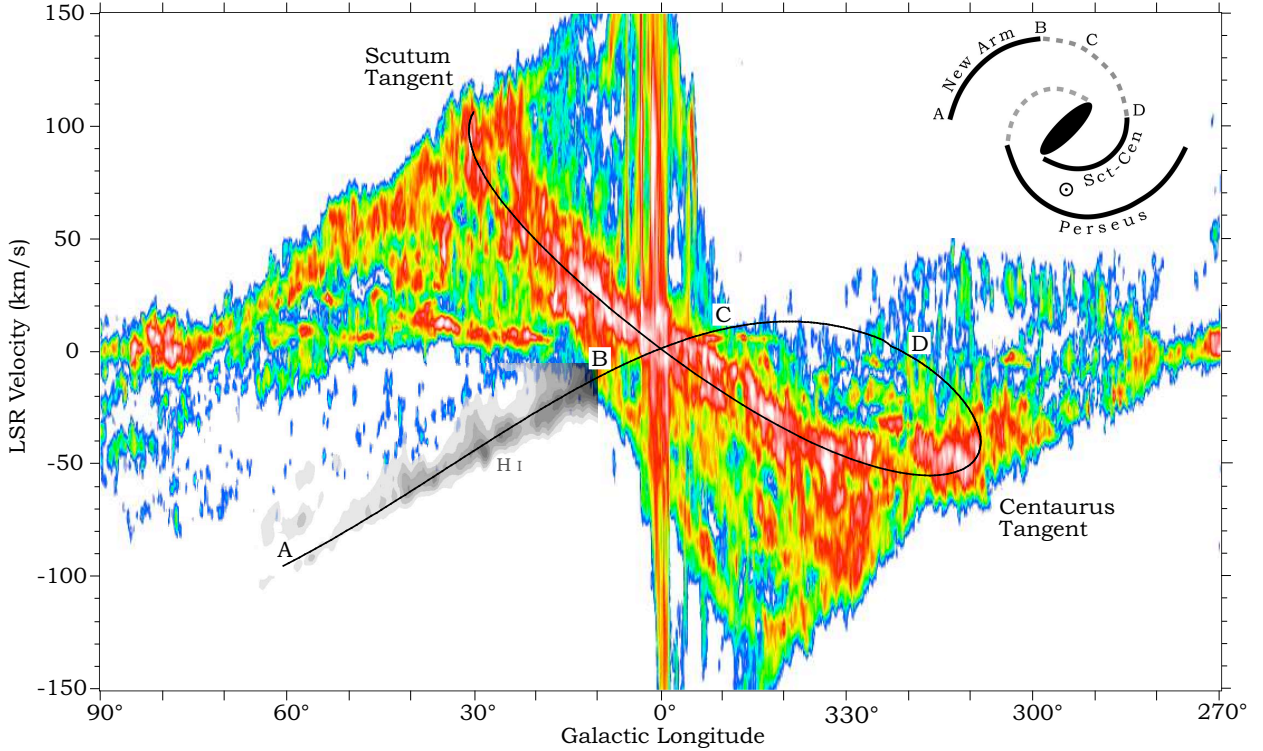}
\caption{ CO longitude-velocity diagram of the first and fourth Galactic quadrants ({\em color}) with the new arm as traced in 21 cm emission overlaid in grayscale (as in Fig. 3b).  CO intensities integrated from $b = -1^{\circ}$ to $1^{\circ}$ are displayed logarithmically from 0.1 K-arcdeg ({\em dark blue}) to 10 K-arcdeg ({\em white}). The black line is the locus of a logarithmic spiral constrained by the tangent directions of the Scutum and Centaurus Arms ({\em labelled}). Specifically, the spiral has tangents at $l = 30.7^{\circ}$ and $l = 308.7^{\circ}$ and a pitch angle of $14.2^{\circ}$. The inset is a schematic face-on view of the Galaxy showing only the two main density-wave arms proposed by Churchwell et al. (2009) extending from opposite ends of the central bar. The dashed sections of the arms have not yet been well traced and so are largely extrapolations. The position of the Sun is indicated. The letters A--D in both the main figure and the inset mark sections of the arm discussed in the text.
 }
\end{figure*}

Confirmation of the present feature as the "Outer Sct-Cen Arm" will require a great deal of new data from several telescopes and much observing time over an extended period. The scale of the problem is evident if we partition the arm into the segments indicated in Figure 4. The present observations of segment AB, covering some 18 kpc in the northern hemisphere is, as noted, severely undersampled and will require several years with the present CfA telescope to flesh out properly. Segment BC, about 7 kpc long where Sct--Cen crosses the nuclear disk and the Galactic center, is of little help to the identification, because it is almost impossible to distinguish it from the mass of unrelated molecular gas in this direction; it may never be adequately deconvolved. The real observational challenge is the 13 kpc segment CD in the fourth Galactic quadrant. Detection and observation of this segment would provide strong support for the present assignment.  

Although segment CD is marginally closer than AB ($\sim$18 kpc), it is much more difficult to track because its velocity hovers so close to zero, where it blends with closer and brighter emission from both local material and the near Carina Arm; this difficulty is even more severe in 21 cm emission owing to its larger velocity dispersion. Identifying star formation in this arm is equally challenging, because it lies behind all the copious star formation in the inner fourth quadrant. Although farther away, the present arm in the first quadrant may be more conducive to such studies, because it extends well above the Galactic plane. 
	 
Our detection of substantial amounts of atomic and molecular gas so far along what is probably the distant end of the Sct--Cen Arm lends some support to the recent proposal that this arm and Perseus are the main density-wave arms of the Milky Way, extending symmetrically from opposite ends of the Galactic bar (Drimmel 2000; Benjamin 2008; Churchwell et al. 2009). If the present arm is in fact the outer end of Sct--Cen, it implies that the full arm extends over 60 kpc and wraps some $310^{\circ}$ around the Galactic center. Key steps toward confirming the proposal include, as mentioned, tracking Sct--Cen in the fourth quadrant and, even harder, tracking the Perseus Arm from the point where it passes inside the solar circle near longitude $50^{\circ}$ to its putative origin at the far end of the bar.  Further study of the Perseus Arm in the third quadrant (e.g., Vazquez et al. 2008) is also called for, since that segment would be the symmetric counterpart of the present arm.  
	 
The Galactic symmetry suggested by the present work and clearly demonstrated by the identification of the Far 3-kpc Arm a few years ago, coupled with evidence for a global two-armed spiral pattern in the old stars, and, indeed, with the discovery of the bar itself, all hint at Galactic spiral structure that is both simpler and more amenable to study than had long been assumed. As emphasized here, much work remains, but aided by greatly improved distances from forthcoming astrometric surveys, a reasonably complete picture of our Galaxy's spiral pattern may be achieved over the next decade.
	 
\acknowledgments 
We are indebted to R. Benjamin and T. Bania for useful discussions and J. Megnia for help with the observations. 
%

%

\clearpage


\begin{thebibliography}{}

\bibitem[Abdo \etal (2010)]{Abdo10}
Abdo, A. A. \etals 2010, \apj, 710, 133

\bibitem[Benjamin(2008)]{Benjamin08}
Benjamin, R. A. 2008, IAU Symposium 254, The Galaxy Disk in Cosmological Context, ed. J. Andersen, J. Bland-Hawthorn, \& B. Nordstr\"{o}m (Cambridge: CUP), 319

\bibitem[Binney \& Merrifield(1998)]{Binney98}
Binney, J., \& Merrifield, M 1998, Galactic Astronomy (Princeton: Princeton Univ. Press)

\bibitem[Churchwell \etal (2009)]{Churchwell09}
Churchwell, E., \etals 2009, \pasp, 121, 213

\bibitem[Dame, Hartmann, \& Thaddeus(2001)]{Dame01}
Dame, T. M., Hartmann, Dap, \& Thaddeus, P. 2001, \apj, 547, 792

\bibitem[Dame \etal (1993)]{Dame93}
Dame, T. M., Koper, E., Israel, F. P., \& Thaddeus, P. 1993, \apj, 418, 730

\bibitem[Dame \& Thaddeus(2008)]{Dame08}
Dame, T. M., \& Thaddeus, P. 2008, \apj, 683, L143

\bibitem[Drimmel(2000)]{Drimmel00}
Drimmel, R. 2000, \aap, 358, L13

\bibitem[Feitzinger \& Spicker(1986)]{Feitzinger86}
Feitzinger, J. V., \& Spicker, J. 1986, \pasj, 38, 485


\bibitem[Kalberla \etal (2005)]{Kalberla05}
Kalberla, P. M. W., Burton, W. B., Hartmann, D., Arnal, E. M., Bajaja, E., Morras, R., \& P\"{o}ppel, W. G. L. 2005, \aap, 440, 775

\bibitem[Levine, Blitz, \& Heiles(2006)]{Levine06}
Levine, E. S., Blitz, L., \& Heiles, C. 2006, \apj, 643, 881

\bibitem[Rolleston \etal (2000)]{Rolleston00}
Rolleston, W. R. J., Smartt, S. J., Dufton, P. L., \& Ryans, R. S. I.  2000, \aap, 363, 537

\bibitem[Stil \etal(2006)]{Stil06}
Stil, J. M., Taylor, A. R., Dickey, J. M., Kavars, D. W., Martin, P. G., Rothwell, T. A., Boothroyd, A. I., Lockman, F. J., \& McClure-Griffiths, N. M.  2006, \aj, 132, 1158

\bibitem[Vazquez \etal (2008)]{Vazquez08}
Vazquez, R. A., May, J., Carraro, G., Bronfman, L., Moitinho, A., \& Baume, G. 2008, \apj, 672, 930

\bibitem[Weaver(1974)]{Weaver74}
Weaver, H. 1974, IAU Symposium 60, Galactic Radio Astronomy, ed. F. J. Kerr \& S. C. Simonson (Dordrecht: Reidel), 573

\end{thebibliography}
\end{document}